\documentclass[aps,pre,reprint,amsmath,amssymb]{revtex4-2}
\usepackage{txfonts}
\usepackage{algorithmic}
\usepackage{algorithm}
\usepackage{graphicx}
\usepackage{bm}
\usepackage{url}
\usepackage{color}
\usepackage{svg}

\newcommand{\argmin}{\mathop{\rm arg~min}\limits}

\begin{document}

\title{Stochastic modeling of deterministic laser chaos using generator extended dynamic mode decomposition}

\author{Kakutaro Fukushi and Jun Ohkubo}

\affiliation{Graduate School of Science and Engineering, Saitama University, Sakura, Saitama, 338--8570 Japan}

\begin{abstract}
Recently, chaotic phenomena in laser dynamics have attracted much attention to its applied aspects, and a synchronization phenomenon, leader-laggard relationship, in time-delay coupled lasers has been used in reinforcement learning. In the present paper, we discuss the possibility of capturing the essential stochasticity of the leader-laggard relationship; in nonlinear science, it is known that coarse-graining allows one to derive stochastic models from deterministic systems. We derive stochastic models with the aid of the Koopman operator approach, and we clarify that the low-pass filtered data is enough to recover the essential features of the original deterministic chaos, such as peak shifts in the distribution of being the leader and a power-law behavior in the distribution of switching-time intervals. We also confirm that the derived stochastic model works well in reinforcement learning tasks, i.e., multi-armed bandit problems, as with the original laser chaos system.
\end{abstract}

\maketitle

\section{Introduction}
\label{sec_introduction}

Coupled lasers have been used as test beds to observe various synchronization phenomena (for example, see \cite{Heil2001,Rogers-Dakin2006,Klein2006,Fisher2006,Vicente2007,Peil2007,Avila2009,Nixon2011,Nixon2012} and the review in \cite{Soriano2013}.) An example of synchronization is the leader-laggard relationship. In the leader-laggard relationship, one of two lasers oscillates in advance of the other by a propagation delay time. The laser oscillating in advance is called the leader, and the other one that oscillates with a delay and follows the leader is called the laggard. Kanno \textit{et al.} reported the spontaneous exchange of the leader-laggard relationship \cite{Kanno2017}; there are autonomous and irregular exchanges between the leader and the laggard.

The fast information processing of lasers could be a candidate for promising key components of artificial intelligence or machine learning. Some works focused on photonic implementations for solving a multi-armed bandit (MAB) problem \cite{Robbins1952}, which is one of the famous examples in reinforcement learning. In the MAB problem, we have several slot machines with unknown hit probabilities. A player should randomly play multiple slot machines at the early stage and estimate the slot machine with the maximum hit probability. This seeking stage is called ``exploration.'' After the exploration, the player selects the slot machine with the maximum hit probability to maximize the total reward obtained from the slot machines. This phase is called ``exploitation.'' Hence, there is a trade-off between the exploration-exploitation in the MAB problem, and it is crucial to strike a balance between the two phases. There are several works on the MAB problem from the viewpoint of photonic implementations (for example, see \cite{Kim2013,Naruse2014,Naruse2015,Naruse2017,Mihana2018,Naruse2018,Ma2020,Iwami2022,Morijiri2023,Cuevas2024,Iwami2024}) and biological ones (for example, see \cite{Kim2010,Kim2015}.) The spontaneous exchange property of the leader-laggard relationship is also desirable for the MAB problem, and the numerical and experimental demonstrations have been reported \cite{Mihana2019,Mihana2020,Ito2024}.

Although the demonstrations for the MAB problem clarified the applicability of lasers, it has not been clear what the necessary components are for this information processing. Since the leader-laggard relationship stems from the lag synchronization of chaos, a deterministic description was employed in the numerical demonstration \cite{Kanno2017,Mihana2019}. On the other hand, one expects that the reinforcement learning tasks need some stochasticity. Then, what aspects of the leader-laggard relationship caused by laser chaos are crucial to the MAB problem? How can we connect deterministic and probabilistic descriptions? As for the connection, one could focus on the fact known in nonlinear science: coarse-graining allows one to derive stochastic models from deterministic systems. One of the theoretical methods for coarse-graining is the Mori-Zwanzig projection method, in which we derive macroscopic equations for relevant variables from microscopic equations \cite{Mori1965,Zwanzig1961,Zwanzig1973}. The irrelevant parts play as noise in the macroscopic equations. The leader-laggard relationship also employs a coarse-graining procedure, i.e., taking the short-term cross-correlations. Hence, it would be valuable to extract the essential parts and obtain stochastic descriptions for the leader-laggard relationship. However, it is difficult to apply the Mori-Zwanzig projection method directly to the leader-laggard relationship because of time-delayed couplings in the time-evolution equations.

\begin{figure}[tb]
\centering
\includegraphics[width=60mm]{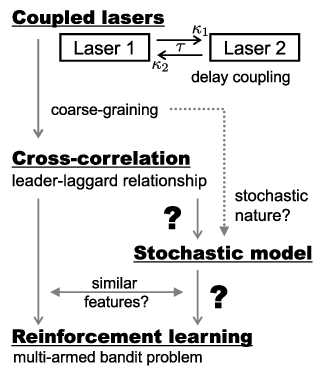}
\caption{
Problem settings. The main aim here is to construct stochastic models from data generated by two lasers mutually coupled with the time-delay $\tau$. The original data stems from deterministic descriptions, and we investigate whether the data generated by the constructed stochastic models has similar features to the original one.
}
\label{fig_concept}
\end{figure}

In the present paper, we aim to construct stochastic models from data generated by two mutually coupled lasers; the coupling has a time-delay, which is crucial to cause the synchronization chaos. Figure~\ref{fig_concept} shows the problem settings of the present paper. As denoted above, the coarse-graining nature in deriving the leader-laggard relationship should guarantee the validity of the probabilistic description, and we must check it. However, as denoted later, it is difficult to construct a stochastic model describing the whole dynamics of the leader-laggard relationship because the dynamics are quite complicated. The key point here is not to perfectly reproduce the behavior, but to reproduce the features that play a crucial role in reinforcement learning. Hence, one should employ appropriate data preprocessing, and the preprocessing contributes to clarifying the essential components of the leader-laggard relationship in the MAB problem. To construct the stochastic models from data, we employ a method in \cite{Tahara2024}, which is based on the Koopman operator approach. The Koopman operator approach enables us to deal with nonlinear systems in terms of linear algebra; we will briefly review the related works in Sec.~\ref{sec_gEDMD}. As for the characteristic statistics for comparison, we focus on two distributions, i.e., the distribution of being the leader and that of switching-time intervals, because they could be essential to achieve reinforcement learning in the MAB problem. After discussing the data preprocessing and features of constructed stochastic models, we finally yield a demonstration for the MAB problem using one of the constructed stochastic models.

The outline of the present paper is as follows. In Sec.~\ref{sec_numerical_models}, we briefly review the coupled lasers and the leader-laggard relationship. Section~\ref{sec_gEDMD} gives a brief review of the Koopman operator approach to constructing stochastic models from data. Section~\ref{sec_main} yields the main contribution of the present paper; we will construct three stochastic models by changing data preprocessing and discuss features of statistics derived by datasets generated by the constructed stochastic models. We also numerically demonstrate the MAB problem. Section~\ref{sec_conclusion} gives some concluding remarks.

\section{Numerical model of laser chaos}
\label{sec_numerical_models}

Here, we briefly explain the numerical model for the coupled semiconductor lasers according to \cite{Kanno2017}. We also describe a method of how the dataset is generated.

\subsection{Time-delayed mutually coupled semiconductor lasers}

\begin{table*}[tb]
\centering
\caption{Parameter values used in numerical simulations}
\label{tab_parameters}
\begin{tabular}{cll}
\hline
Symbol & Parameter & Value \\
\hline
$G_N$ & Gain coefficient & $8.40 \times 10^{-13} \, \mathrm{m}^{3} \mathrm{s}^{-1}$\\
$N_0$ & Carrier density at transparency & $1.40 \times 10^{24} \,\mathrm{m}^{-3}$  \\
$\epsilon$ & Gain saturation coefficient & $4.5 \times 10^{-23}$\\
$\tau_\mathrm{p}$ & Photon lifetime & $1.927 \times 10^{-12} \, \mathrm{s}$\\
$\tau_\mathrm{s}$ & Carrier lifetime & $2.04 \times 10^{-9} \, \mathrm{s}$\\
$\alpha$ & Linewidth enhancement factor & $3.0$\\
$\lambda_1$ & Optical wavelength of Laser 1 & $1.537 \times 10^{-6} \, \mathrm{m}$\\
$c$ & Speed of light & $2.998 \times 10^{8} \, \mathrm{m}\mathrm{s}^{-1}$\\ 
$\kappa_1$ & Coupling strength from Laser 1 to Laser 2 & $30.00 \times 10^{-9} \, \mathrm{s}^{-1}$ (Variable)\\
$\kappa_2$ & Coupling strength from Laser 2 to Laser 1 & $30.00 \times 10^{-9} \, \mathrm{s}^{-1}$ (Variable)\\
\begin{minipage}{25mm}
\centering
\vspace{1mm}
$N_\mathrm{th}$\\
{\footnotesize $(= N_0+ 1/(G_N \tau_\mathrm{p}))$}
\end{minipage}
 & Carrier density at lasing threshold & $2.018 \times 10^{24} \, \mathrm{m}^{-3}$\\
\begin{minipage}{25mm}
\centering
\vspace{1mm}
$J_\mathrm{th}$\\
{\footnotesize $(= N_\mathrm{th} / \tau_\mathrm{s})$}
\end{minipage}
& Injection current at lasing threshold & $9.892 \times 10^{32} \, \mathrm{m}^{-3} \mathrm{s}^{-1}$ \\
$J$ & Injection current & $1.1 J_\mathrm{th}$ \\
$\tau$ & Propagation delay time of light between two lasers & $36.64 \times 10^{-9} \, \mathrm{s}$\\
$\Delta f_{\mathrm{ini}}$ & Initial optical frequency detuning between two lasers & $1.0 \times 10^{9} \, \mathrm{Hz}$ (Variable)\\
\hline
\end{tabular}
\end{table*}

The model of mutually coupled semiconductor lasers is depicted at the top of Figure~\ref{fig_concept}; two semiconductor lasers are mutually coupled with a coupling delay time $\tau$. Note that the coupling strengths $\kappa_1$ and $\kappa_2$ could be varied to control the leader-laggard relationship, as denoted later. The time-evolution equations of the numerical model for mutually coupled semiconductor lasers are described by the Lang-Kobayashi equations \cite{Lang1980,Kanno2017}. Let $E_n$ be the complex electric-field amplitude of the $n$-th laser, and the carrier density of the $n$-th laser is denoted as $N_n$. Then, the Lang-Kobayashi equations are described as follows:
\begin{align}
\frac{dE_{n}(t)}{dt} &= \frac{1+i\alpha}{2}\left[\frac{G_{N}\left(N_{n}(t)-N_0\right)}{1+\epsilon\left|E_{n}(t)\right|^2}-\frac{1}{\tau_p}\right]E_{n}(t) \nonumber \\
&\qquad+ \kappa_{n+1} E_{n+1}(t-\tau)e^{i\theta_{n}(t)}, \label{eq_LK_1}\\
\frac{dN_{n}(t)}{dt} &= J-\frac{N_{n}(t)}{\tau_s}-\frac{G_N\left(N_{n}(t)-N_0\right)}{1+\epsilon\left|E_{n}(t)\right|^2}\left|E_{n}(t)\right|^2, \label{eq_LK_2}\\
\theta_{n}(t) &= (\omega_{n+1}-\omega_{n})t-\omega_{n+1}\tau, \label{eq_LK_3}
\end{align}
where $n \in \{1,2\}$ corresponds to the $n$-th laser, and $n=3$ means Laser 1, i.e., $E_3(t) \equiv E_1(t)$ and $N_3(t) \equiv N_1(t)$. The symbol $\theta_n(t)$ is the optical phase difference between the laser light and the injected light, and $\omega_1 = 2\pi c / \lambda_1$ is the optical angular frequency for the first laser. Note that $\Delta\omega = 2\pi\Delta f_\textrm{ini}$ and $\omega_2 = \omega_1 - \Delta \omega$. The symbol $\Delta f_\textrm{ini}$ represents the initial optical frequency detuning between the two lasers; $\Delta f_\textrm{ini}$ is an important variable to control the characteristics of the leader-laggard relationship. The coupling strengths $\kappa_1$ and $\kappa_2$ are also crucial, and we will change them in the numerical experiments described later. The meanings of other symbols and their values are summarized in Table~\ref{tab_parameters}. Here, we basically employ the same settings with \cite{Kanno2017}.

\begin{figure}[bt]
\centering
\includegraphics[width=70mm]{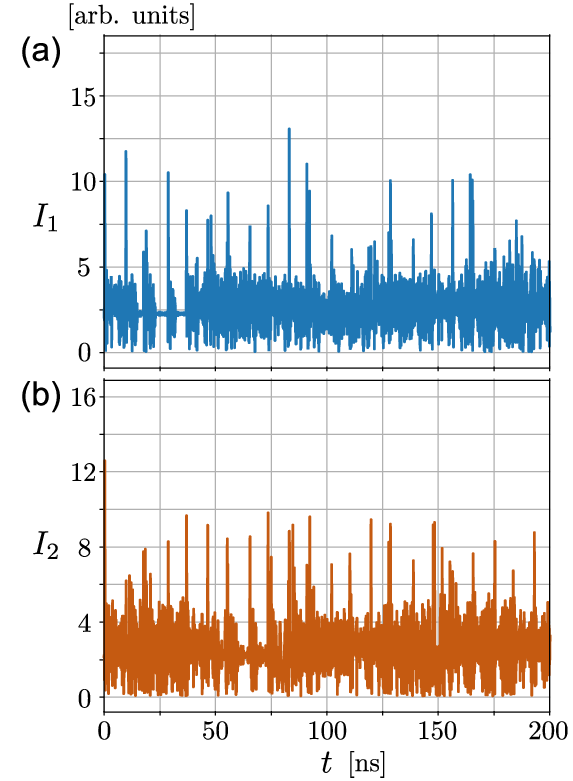}
\caption{Sample trajectories of optical intensities for Laser 1 (a) and Laser 2 (b). The vertical axis is drawn using arbitrary unit. The initial optical frequency detuning $\Delta f_\mathrm{ini}$ is set to $1$ GHz.
}
\label{fig_intensity}
\end{figure}

We numerically integrate the Lang-Kobayashi equations in Eqs.~\eqref{eq_LK_1}, \eqref{eq_LK_2} and \eqref{eq_LK_3} to generate the original data for the chaotic laser dynamics. Here, the fourth order Runge-Kutta method with the time-interval $\Delta t = 1.0\times 10^{-12} \, \mathrm{s}$ is employed. 

Figure~\ref{fig_intensity} shows examples of the intensities of the $n$-th laser, $I_n(t)$, where $I_n(t) = \lvert E_n(t) \rvert^2$. Since plots with $\Delta t = 1.0\times 10^{-12} \, \mathrm{s}$ are too dense, we plotted the trajectories with the time-interval $1.0\times 10^{-10} \, \mathrm{s}$. Note that the behavior of Laser 1 (Fig.~\ref{fig_intensity}(a)) around $t=25$ is similar to that of Laser 2 (Fig.~\ref{fig_intensity}(b)) around $61 (\simeq 25 + \tau)$. Hence, there would be a synchronization of the laser chaos. Thus, we see that the intensity of Laser 2 would lag behind that of Laser 1 by the propagation delay time $\tau$.

\subsection{Cross-correlation and leader-laggard relationship}

\begin{figure}[bt]
\centering
\includegraphics[width=80mm]{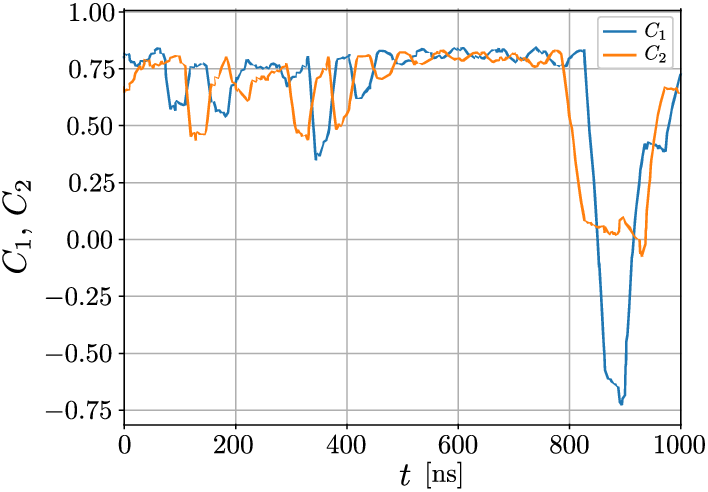}
\caption{Examples of trajectories of short-term cross-correlation values $C_1(t)$ and $C_2(t)$. These cross-correlation values are evaluated from the optical intensities generated by the same settings as Fig.~\ref{fig_intensity}.
}
\label{fig_cross_correlation}
\end{figure}

For applications in machine learning of coupled lasers, short-term correlations are used rather than original trajectories in Fig.~\ref{fig_intensity}. The short-term cross-correlation values for Laser 1, $C_1(t)$, and those for Laser 2, $C_2(t)$, are given as follows \cite{Kanno2017}:
\begin{align}
C_1(t)=\frac{\left\langle\left[I_1(t-\tau)-\overline{I}_1\right]\left[I_2(t)-\overline{I}_2\right]\right\rangle_\tau}{\sigma_1\sigma_2},\\
C_2(t)=\frac{\left\langle\left[I_1(t)-\overline{I}_1\right]\left[I_2(t-\tau)-\overline{I}_2\right]\right\rangle_\tau}{\sigma_1\sigma_2},
\end{align}
where $\overline{I}_n$ and $\sigma_n$ are the mean and the standard deviation values of the temporal intensity for the $n$-th laser for the period $\tau$. $\langle\cdot\rangle_\tau$ represents the time average over the period $\tau$. 

The comparison of the cross-correlation values $C_1$ and $C_2$ leads to the leader-laggard relationship. That is, when $C_1 > C_2$, one considers that $I_1(t-\tau)$ and the time-delayed value $I_2(t)$ show large synchronization. This indicates that Laser 1 is the leader. Conversely, $C_2 > C_1$ means that Laser 2 is the leader. Hence, comparing $C_1(t)$ with $C_2(t)$ immediately helps us to determine the leader-laggard relationship.

Figure~\ref{fig_cross_correlation} shows examples of the evaluated short-term cross-correlation values. The experimental setting is the same as in Fig.~\ref{fig_intensity}; we can see the following points \cite{Kanno2017}:
\begin{itemize}
\item There are periodic switchings of the leader-laggard relationship with the period $\tau$.
\item The periodic switching behavior is not completely regular but has a randomness.
\item The trajectories of the short-term cross-correlations could show sudden dropout (around $t=800$ [ns] in Fig.~\ref{fig_cross_correlation}) and the subsequent gradual recovery.
\end{itemize}

As denoted in Sec.~\ref{sec_introduction}, one could see a kind of stochasticity in the behavior of the short-term cross-correlations; we will see some distributions evaluated by the trajectories of $C_1(t)$ and $C_2(t)$ in Sec.~\ref{sec_main}. Although it might be difficult to see in Fig.~\ref{fig_cross_correlation}, $C_1(t)$ is likely larger than $C_2(t)$ in this experimental setting. Some parameters can adjust the tendency of which laser is more likely to be the leader. In \cite{Kanno2017}, the tuning parameter is the initial optical frequency detuning between the two lasers, $\Delta f_{\mathrm{ini}}$; in \cite{Mihana2019}, the difference in the coupling strengths, $\kappa_1 - \kappa_2$ yields the control of the probability of becoming the leader laser. Note that the reinforcement learning for the MAB problem exploited the features of the short-term cross-correlations \cite{Mihana2019}. That is, although the original trajectories for the intensity in Fig.~\ref{fig_intensity} are too complicated, the trajectories have undergone a kind of coarse-graining, and they contain some stochasticity. Hence, we will consider stochastic models for the short-term cross-correlations below.

\section{Equation estimation via the Koopman generator}
\label{sec_gEDMD}

Here, we yield a brief review of a method developed in \cite{Tahara2024} to estimate stochastic differential equations from data. 

\subsection{Koopman operator theory}

The Koopman operator theory \cite{Koopman1931} has a long history, and the recent advances in data-driven science have expanded further applications of the Koopman theory. The famous data-driven approach is the dynamic mode decomposition (DMD) \cite{Schmid2010}, which enables us to extract information about the Koopman operator from time-series data. An extension of the DMD called the Hankel alternative view of Koopman (HAVOK) was applied to investigate the characteristics of chaotic systems \cite{Brunton2017}. Another extension utilizing basis functions is the extended dynamic mode decomposition (EDMD) \cite{Williams2015}. As for the details of the Koopman theory, see the review papers \cite{Mezic2021} and \cite{Brunton2022}. Here, we focus only on the framework of equation estimation based on the Koopman operator theory.

One of the well-known methods of equation estimation is the sparse identification of non-linear dynamics (SINDy) \cite{Brunton2016}, which utilizes sparse estimation techniques to infer equations from data for deterministic dynamical systems. Although there are some extensions of the SINDy, for example, see \cite{Boninsegna2018}, the Koopman operator theory yields a straightforward method to estimate the underlying equations for a dataset; note that the underlying equations could be not only deterministic but also stochastic. The generator extended dynamic mode decomposition (gEDMD) is an extension of the EDMD algorithm for estimating the underlying equations \cite{Klus2020}. There are some applications of the gEDMD algorithm, such as model reduction \cite{Niemann2021} and prediction \cite{Klus2020-2}. However, the estimation is difficult for the stochastic cases because of noise in the dataset. In \cite{Tahara2024}, several data preprocessing methods are proposed, which yield reasonable estimations of the underlying stochastic differential equations from an artificially generated dataset. Since the method in \cite{Tahara2024} works well even in a noisy dataset, we employ it to construct stochastic models for the short-term cross-correlations in Fig.~\ref{fig_cross_correlation}.

\subsection{Brief summary of gEDMD}

We here describe only the actual estimation procedure; for the details, see \cite{Tahara2024}.

Our final aim is to construct the following stochastic diffusion equation for a state vector $\bm{X}(t) \in \mathbb{R}^D$ \cite{Gardiner_book}:
\begin{align}
d\bm{X}(t) = \bm{b}(\bm{X}(t))dt + \sigma(\bm{X}(t)) d\bm{W}(t),
\label{eq_sde}
\end{align}
where the function $\bm{b}(\bm{x}): \mathbb{R}^{D}\rightarrow\mathbb{R}^{D}$ is a $D$-dimensional vector function which is referred as the drift coefficients, and $\sigma(\bm{x}): \mathbb{R}^{D}\rightarrow\mathbb{R}^{D\times D}$ is a matrix function which is referred as the diffusion coefficients. $W(t)$ is a $D$-dimensional Wiener process. In the short-term cross-correlation cases, we have two variables $C_1(t)$ and $C_2(t)$, and hence $D=2$. Then, our task is to estimate the function forms $\bm{b}(\bm{x})$ and $\sigma(\bm{x})$.

Here, we introduce an infinitesimal generator $\mathcal{L}$, a dictionary, and a least-squares problem. Given a twice continuously differentiable function $f$, the infinitesimal generator $\mathcal{L}$, which is deeply related to the Koopman operator, is defined as follows:
\begin{align}
(\mathcal{L}f)(\bm{x})
&=\sum_{i=1}^{D} b_i(\bm{x})\frac{\partial}{\partial x_i}f(\bm{x})+\frac{1}{2}\sum_{i=1}^{D}\sum_{j=1}^{D}a_{ij}(\bm{x})\frac{\partial^2}{\partial x_i\partial x_j}f(\bm{x}),
\label{eq_generator}
\end{align}
where $A(\bm{x}) = \sigma(\bm{x})\sigma(\bm{x})^\top$, and $a_{ij}(\bm{x})$ is the $ij$ component of the matrix $A(\bm{x})$.

A set of basis functions, $\{\psi_k(\bm{x})\}_{k=1}^{K}$, is called a dictionary. Since monomial basis functions are easy to treat, we here employ them as in \cite{Tahara2024}. For the two-dimensional case, the dictionary is set as follows:
\begin{align}
\bm{\psi}(\bm{x}) &= [\psi_1(\bm{x}),\psi_2(\bm{x}),\psi_3(\bm{x}),\psi_4(\bm{x}),\psi_5(\bm{x}),\dots]^{\top} \nonumber \\
&= [1, x_1, x_2, x_1^2, x_1 x_2, \dots]^{\top}.
\label{eq_basis_function}
\end{align}
The action of the generator $\mathcal{L}$ on the basis function on a specific data point $\bm{x}_\ell$ is then described by
\begin{align}
&d\psi_k(\bm{x}_\ell) \nonumber \\
&\equiv 
(\mathcal{L}\psi_k)(\bm{x}_\ell) \nonumber \\
&=
\sum_{i=1}^{D} b_i(\bm{x}_\ell)\frac{\partial}{\partial x_i}\psi_k(\bm{x}_\ell)+\frac{1}{2}\sum_{i=1}^{D}\sum_{j=1}^{D}a_{ij}(\bm{x}_\ell)\frac{\partial^2}{\partial x_i\partial x_j}\psi_k(\bm{x}_\ell).
\label{eq_generator_data}
\end{align}
Note that the functions $\{b_i(\bm{x})\}$ and $\{a_{ij}(\bm{x})\}$ are unknown. However, it is possible to estimate the concrete values on a specific data point, $\{b_i(\bm{x}_\ell)\}$ and $\{a_{ij}(\bm{x}_\ell)\}$, in Eq.~\eqref{eq_generator_data} from the data. As discussed in \cite{Klus2020}, these functions are defined as the following conditional expectations:
\begin{align}
\bm{b}(\bm{x}_\ell)&= \lim_{t\to 0}\mathbb{E}\left[\left. \frac{1}{t}(\bm{X}(t)-\bm{x}_\ell) \right| \bm{X}(0)=\bm{x}_\ell\right], \label{eq_conditional_expectation_b}\\
A(\bm{x}_\ell)&= \lim_{t\to 0}\mathbb{E}\left[\left. \frac{1}{t}(\bm{X}(t)-\bm{x}_\ell)(X(t)-\bm{x}_\ell)^\top \right|\bm{X}(0)=\bm{x}_\ell\right].
\label{eq_conditional_expectation_A}
\end{align}

In \cite{Tahara2024}, the conditional expectations in Eqs.~\eqref{eq_conditional_expectation_b} and \eqref{eq_conditional_expectation_A} are evaluated via weighted averages. Here, assume that we have a trajectory dataset $\bm{x}(t)$ as in Fig.~\ref{fig_cross_correlation}, and we prepare a total of $N$ data points $\{\bm{x}_n\}$ with the time-interval $\Delta t$, i.e., $\bm{x}_1 = \bm{x}(0), \bm{x}_2 = \bm{x}(\Delta t), \dots, \bm{x}_N = \bm{x}((N-1)\Delta t)$. However, due to the presence of noise in the dataset, the estimation of the conditional expectations needs large amounts of data. As explained later, the final least-squares problem is difficult to solve when the data size is large, and then the following two data-preprocessing methods were introduced in \cite{Tahara2024}:
\begin{itemize}
\item Clustering for selecting a small number of representative points $\{\bm{x}_\ell\}_{\ell=1}^{N_{\mathrm{r}}}$. The $k$-means clustering algorithm is employed here.
\item Clustering for evaluating covariance matrices depending on coordinates. The clustering method based on the Dirichlet Process Mixture Model (DPMM)\cite{Ferguson1973} is employed here.
\end{itemize}
We perform the $k$-means clustering algorithm to select $N_\mathrm{r}$ representative points in the first clustering method. Note that $N_\mathrm{r}$ is considerably smaller than the original data size $N$, which is preferable to solve the least-squares problem. In \cite{Tahara2024}, it was shown that the small number of representative points is enough to estimate the underlying equations if one adequately evaluates the conditional expectations in Eqs.~\eqref{eq_conditional_expectation_b} and \eqref{eq_conditional_expectation_A}. For the accurate estimations of $\bm{b}(\bm{x}_\ell)$ and $A(\bm{x}_\ell)$ on a representative point $\bm{x}_\ell$, the second clustering method was employed, in which $N_\mathrm{c}$ clusters are generated; the number of clusters, $N_\mathrm{c}$, is determined automatically by the DPMM. Note that $N_\mathrm{c}$ is considerably smaller than $N_\mathrm{r}$, and we use the generated clusters to evaluate the covariance matrices depending on coordinates. Let $\mathcal{C}_p$ be the set of indices of the original dataset for the $p$-th cluster, and assume that a representative point $\bm{x}_\ell$ belongs to the $p$-th cluster. Then, the conditional expectations on $\bm{x}_\ell$ are evaluated via the following weighted averages:
\begin{align}
\widetilde{\bm{b}}(\bm{x}_\ell)&=\frac{1}{Z_{\bm{x}_\ell}} \sum_{n \in \mathcal{C}_p} K_{\bm{H}_p}(\bm{x}_\ell,\bm{x}_n)\left[\frac{1}{\Delta t}(\bm{x}_{n+1}-\bm{x}_{n})\right],\\
\widetilde{A}(\bm{x}_\ell)&=\frac{1}{Z_{\bm{x}}}\sum_{n \in \mathcal{C}_p} K_{\bm{H}_p}(\bm{x}_\ell,\bm{x}_n)\left[\frac{1}{\Delta t}(\bm{x}_{n+1}-\bm{x}_{n})(\bm{x}_{n+1}-\bm{x}_{n})^\top\right],\\
Z_{\bm{x}_\ell} &= \sum_{n \in \mathcal{C}_p}K_{\bm{H}_p}(\bm{x}_\ell,\bm{x}_n),
\end{align}
where $K_{\bm{H}_p}(\bm{x},\bm{x}')$ is a kernel function for the $p$-th cluster. Here, we use the Gaussian kernel:
\begin{align}
K_{\bm{H}_p}(\bm{x},\bm{x}')=\mathrm{exp}\left(-\frac{1}{2}(\bm{x}'-\bm{x})^\top H_p^{-1}(\bm{x}'-\bm{x})\right),
\label{eq:kernel}
\end{align}
where $H_p$ represents the bandwidth matrix; based on empirical rules, $H_p$ is given as follows:
\begin{align}
H_p = \left(\beta N_p^{-\frac{1}{D+2}} \right)^2 \Sigma_p,
\label{eq:band}
\end{align}
where $\beta$ is a hyperparameter, $N_p$ is the number of points in the $p$-th cluster, and $\Sigma_p$ represents the covariance matrix for the $p$-th cluster.

Let $\bm{d\psi}(\bm{x}_\ell)$ be the following vector:
\begin{align}
\bm{d\psi}(\bm{x}_\ell) = [d\psi_1(\bm{x}_\ell),d\psi_2(\bm{x}_\ell),\dots, d\psi_K(\bm{x}_\ell)]^\top.
\end{align}
Then, we have the following least-squares problem:
\begin{align}
M = \argmin_{\widetilde{M}} \sum_{\ell=1}^{N_\mathrm{r}} \left\| d\psi(\bm{x}_\ell) -  \widetilde{M} \psi(\bm{x}_\ell)\right\|.
\label{eq_least_squares}
\end{align}
The estimated matrix $M$ is called the Koopman generator matrix \cite{Klus2020}. In the practical numerical estimation in Sec.~\ref{sec_main}, the least-squares problem will be solved with a regularization term for sparsity; we will employ the least absolute shrinkage and selection operator (LASSO) algorithm with a hyperparameter $\lambda$.

As the final step of the gEDMD algorithm, we connect the Koopman generator matrix with the equation estimation problem. The following example of the two-dimensional case makes it easier to understand the estimation of $b_1(\bm{x})$; the second basis function is $\psi_2(\bm{x}) = x_1$ in Eq.~\eqref{eq_basis_function}, and the action of the generator $\mathcal{L}$ on $\psi_2(\bm{x})$ leads to 
\begin{align}
(\mathcal{L}\psi_2)(\bm{x})
&=
\sum_{i=1}^{2} b_i(\bm{x})\frac{\partial}{\partial x_i} x_1 +\frac{1}{2}\sum_{i=1}^{2}\sum_{j=1}^{2}a_{ij}(\bm{x})\frac{\partial^2}{\partial x_i\partial x_j} x_1  \nonumber \\
&= b_{1}(\bm{x}).
\end{align}
Note that $\mathcal{L} \psi_2(\bm{x}) = d\psi_2(\bm{x})$, and $d\psi_2(\bm{x})$ is approximated as the second row of $M \bm{\psi}(\bm{x})$ according to Eq.~\eqref{eq_least_squares}. Hence, we have
\begin{align}
b_{1}(\bm{x}) \simeq \sum_{k} M_{2k} \psi_k(\bm{x}).
\end{align}
Similarly, $\psi_3(\bm{x}) = x_2$ leads to 
\begin{align}
b_{2}(\bm{x}) \simeq \sum_{k} M_{3k} \psi_k(\bm{x}).
\end{align}
Defining $\psi_4(\bm{x}) = x_1^2$, $\psi_5(\bm{x}) = x_1 x_2$, and $\psi_6(\bm{x}) = x_2^2$, the similar discussions yield the following estimation results for the diffusion coefficients:
\begin{align}
a_{11}(\bm{x}) &\simeq \sum_{k} M_{4k} \psi_k(\bm{x}) - 2 x_1 \sum_{k} M_{2k} \psi_k(\bm{x}), \\
a_{12}(\bm{x}) &= a_{21}  \nonumber \\
&\simeq \sum_{k} M_{5k} \psi_k(\bm{x}) - x_2 \sum_{k} M_{2k} \psi_k(\bm{x})
 - x_1 \sum_{k} M_{3k} \psi_k(\bm{x}), \\
a_{22}(\bm{x}) &\simeq \sum_{k} M_{6k} \psi_k(\bm{x}) - 2 x_2 \sum_{k} M_{3k} \psi_k(\bm{x}).
\end{align}
Finally, the Cholesky decomposition of $A(\bm{x})$ yields the diffusion coefficients $\sigma(\bm{x})$ in Eq.~\eqref{eq_sde}; note that $\sigma(\bm{x})$ is not uniquely determined because of the non-uniqueness of the Cholesky decomposition. Here, we employ \verb|scipy.linalg.sqrtm| \cite{scipy} instead of the Cholesky decomposition; after calculating the matrix square root of $A(\bm{x})$ with \verb|scipy.linalg.sqrtm|, we set its real part as $\sigma(\bm{x})$.

\section{Construction and analysis of stochastic models for cross-correlations}
\label{sec_main}

\subsection{Data preparation}

As in Sec.~\ref{sec_numerical_models}, we firstly prepare the original time-trajectories for the intensities $I_1(t)$ and $I_2(t)$. The Lang-Kobayashi equations in Eqs.~\eqref{eq_LK_1}, \eqref{eq_LK_2}, and \eqref{eq_LK_3} are solved using the fourth order Runge-Kutta method with the time-interval $\Delta t = 1.0\times 10^{-12} \, \mathrm{s}$. After several initial relaxation steps, we perform the time-evolution with $100 \times \tau$, and the short-term cross-correlations $C_1(t)$ and $C_2(t)$ are evaluated. Note that the time-interval $\Delta t = 1.0\times 10^{-12} \, \mathrm{s}$ is too short for the data analysis, and hence we take data points with the time-interval $\Delta t_{\mathrm{obs}} = 1.0\times 10^{-10} \, \mathrm{s}$. Finally, we take $N=300,000$ data points, i.e., $\bm{x}_n = [C_1(n \Delta t_{\mathrm{obs}}),C_2(n \Delta t_{\mathrm{obs}})]^\top$.

In the gEDMD algorithm, the dictionary $\bm{\psi}(\bm{x})$ consists of the monomial basis functions with the maximum degree of 10. Then, the total number of dictionary functions is $66$. We set $N_\mathrm{r} = 100$. The hyperparameter for the bandwidth matrix is set to $\beta = 1.0$.

\subsection{Filtered trajectories}

\begin{figure}[tb]
\centering
\includegraphics[width=80mm]{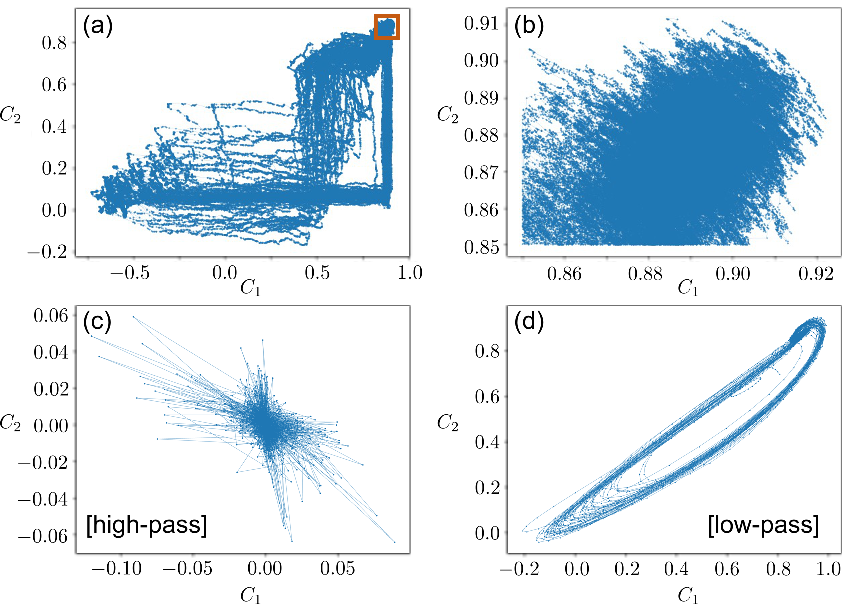}
\caption{Training trajectory data for the gEDMD algorithm. (a) Original trajectory. (b) Extracted dataset within the high-density cluster region. (c) Preprocessed dataset by the high-pass filter with $5.0 \times 10^7$ Hz. (d) Preprocessed dataset by the low-pass filter with $5.0 \times 10^6$ Hz. The orange-colored rectangle in (a) indicates the high-density cluster region depicted in (b).}
\label{fig_training}
\end{figure}

Figure~\ref{fig_training}(a) shows the original trajectory data for $C_1(t)$ and $C_2(t)$. This trajectory data shows the following two characteristics:
\begin{itemize}
\item There is a dense region for large values of $C_1(t)$ and $C_2(t)$, in which most data points are concentrated in this region.
\item Occasionally, it shows the trajectory of a large movement toward negative values of $C_1(t)$ and $C_2(t)$, which corresponds to a sudden dropout in Fig.~\ref{fig_cross_correlation}.
\end{itemize}
These complicated behaviors prevent us from constructing the corresponding stochastic model. In fact, the application of the gEDMD algorithm cannot yield a meaningful model. 

Hence, we consider the effects of the following three data preprocessing procedures.
\begin{itemize} 
\item The preprocessing ``high-density cluster region'' focuses on reproducing only the region $C_1, C_2 > 0.85$ where we have a concentration of most of the data points. This preprocessing aims to capture only the data features within the high-density region. Note that the overall trajectory is not learned. 
\item The preprocessing ``filtering with the high-pass filter'' focuses on reproducing the exchange of the leader-laggard relationship by filtering out the low-frequency components of the data. This approach is intended to capture the dynamics associated with the noisy movement in the original trajectory.
\item The preprocessing ``filtering with the low-pass filter'' focuses on reproducing the probability of being the leader by filtering out the high-frequency components of the data. This approach is intended to capture the dynamics associated with the probability of being the leader. 
\end{itemize}
Note that the high-pass and low-pass filters are applied to the cross-correlation trajectory dataset in Fig.~\ref{fig_training}(a). In \cite{Kanno2017}, the filter was applied to the intensity of the original laser chaos, as in Fig.~\ref{fig_intensity}.

Figures~\ref{fig_training}(b), \ref{fig_training}(c), and \ref{fig_training}(d) show the preprocessed trajectories. After the preprocessing, we construct stochastic models for the preprocessed data in Figs~\ref{fig_training}(b), \ref{fig_training}(c), and \ref{fig_training}(d), respectively. For the LASSO algorithm, we used the hyperparameter $\lambda = 3.0 \times 10^{-9}$, $3.0 \times 10^{-6}$, and $5.0 \times 10^{-7}$, for the trajectories in Figs~\ref{fig_training}(b), \ref{fig_training}(c), and \ref{fig_training}(d), respectively. Note that we tried several hyperparameter values for each case and selected parameters to yield reasonable stochastic models.

\subsection{Features of statistics}

\begin{figure}[tb]
\centering
\includegraphics[width=80mm]{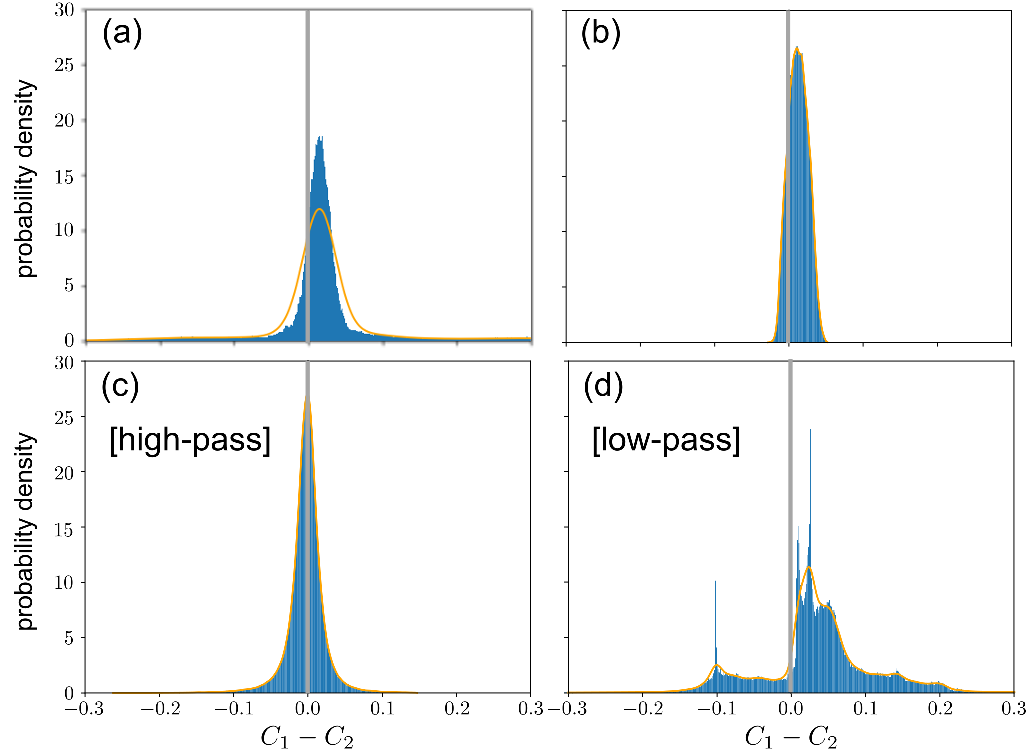}
\caption{The distribution of being the leader. The distribution evaluated by the original trajectory is (a). By contrast, (b), (c), and (d) are evaluated by the artificial dataset generated by the stochastic models constructed from the trajectories in Figs.~\ref{fig_training}(b), \ref{fig_training}(c), and \ref{fig_training}(d), respectively. The blue lines correspond to the histograms, and the orange curves are obtained from the kernel density estimation. The vertical gray lines correspond to $C_1-C_2 = 0.0$.
}
\label{fig_dist1}
\end{figure}

\begin{figure*}[tb]
\centering
\includegraphics[width=120mm]{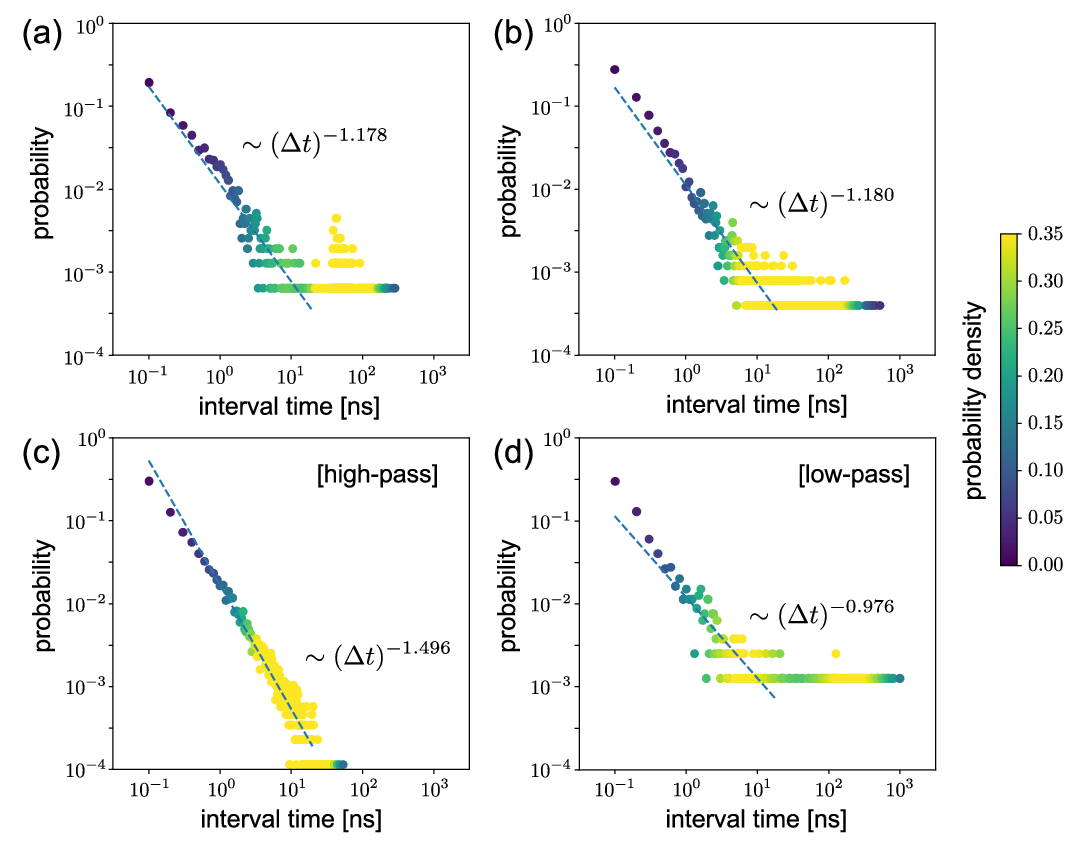}
\caption{The distribution of leader switching-time intervals from Laser 1 to Laser 2. The distribution evaluated by the original trajectory is (a). By contrast, (b), (c), and (d) are evaluated by the artificial dataset generated by the stochastic models constructed from the trajectories in Figs.~\ref{fig_training}(b), \ref{fig_training}(c), and \ref{fig_training}(d), respectively. The color represents the density of non-zero elements in the histograms, which is evaluated using a kernel-density estimation method. The dashed lines correspond to fitting results with power-law decays.
}
\label{fig_dist2}
\end{figure*}

After the construction of the stochastic models, we generate artificial datasets using the constructed models. Monte Carlo simulations with the Euler-Maruyama scheme are used for the time-evolution of the stochastic differential equations; the time-interval for the Euler-Maruyama scheme is $\Delta t = 1.0 \times 10^{-10} \, \mathrm{s}$. In the Monte Carlo simulations, there were exceptional cases where the coordinates tended to diverge. Hence, we employed the following experimental setting: When the simulated trajectory deviated from the region within the original data, we regenerated the noise term; if the regeneration procedure was repeated 10 times, we restarted the time-evolution with randomly selected initial coordinates. Using these generated data points, we evaluate distributions of the following two statistics:
\begin{itemize}
\item The distribution of being the leader. Here, we evaluate the number of data points with $C_1(t) > C_2(t)$, i.e., the number of times that Laser 1 becomes the leader.
\item The distribution of leader switching-time intervals from Laser 1 to Laser 2, i.e., the time-interval taken for $C_1(t) > C_2(t)$ to become $C_2(t) < C_1(t)$.
\end{itemize}

The distribution of being the leader is crucial for achieving reinforcement learning for the MAB problem. Figure~\ref{fig_dist1}(a) shows the distribution of being the leader for the original trajectory in Fig.~\ref{fig_training}(a), in which we can see a positive peak shift. In the figures, we depict not only the histograms, but also the curves which are obtained from the kernel density estimation with \texttt{scipy.stats.gaussian\_kde}; the bandwidth is determined by the Scott's rule $n^{(-1/(d+4))}$, where $n$ is the number of data points and $d$ is the number of dimensions. In the numerical simulation, Laser 1 tends to be the leader because we used $\Delta f_\mathrm{ini} = 1.0 \times 10^9$ Hz. Note that a negative $\Delta f_\mathrm{ini}$ leads to the peak shift toward a negative direction. In addition, larger values of $\Delta f_\mathrm{ini}$ cause larger peak shifts. These characteristics have already been reported in \cite{Kanno2017}. In \cite{Mihana2019}, this feature of the peak shift from control parameters is exploited to solve the reinforcement learning tasks, where $\kappa_1$ and $\kappa_2$ were used as the control parameters instead of $\Delta f_\mathrm{ini}$.

The distributions evaluated by artificial datasets from the stochastic models are shown in Figs.~\ref{fig_dist1}(b), \ref{fig_dist1}(c), and \ref{fig_dist1}(d). Since we will employ the dependency between the peak in $C_1-C_2$ and the initial frequency difference $\Delta f_{\mathrm{ini}}$ in the MAB problem, as denoted above, we here focus on the peak shifts in Figs.~\ref{fig_dist1}(b) and \ref{fig_dist1}(d). Then, there is no peak shift in Fig.~\ref{fig_dist1}(c), and hence, the high-pass filtered data cannot preserve this essential characteristic. The peak shift needs a bias on the state space on $C_1(t)$ and $C_2(t)$; in Figs.~\ref{fig_training}(a), \ref{fig_training}(b), and \ref{fig_training}(d), we see a bias of the data toward the lower right on the diagonal line. These facts indicate that high-frequency noise focused on the high-pass filtered data is not crucial in recovering the leader-laggard relationship. Here, we comment on the complicated shape in Fig.~\ref{fig_dist1}(d). The main focus here is not on the complete recovery of statistical properties. Although the preprocessing with the low-pass filter yields a complicated shape, it is crucial to investigate whether the preprocessing preserves the essential features for the MAB problem. We will discuss this point in Sec.~IV.D.

Figure~\ref{fig_dist2} shows the distributions of leader switching-time intervals from Laser 1 to Laser 2. Here, the leader switching-time interval means the time-interval at which there are switching from $C_1-C_2 \ge 0$ to $C_1-C_2 < 0$. First, we construct the histogram from the simulated data; the bin width is set to $1.0\times 10^{-10}$ s, which corresponds to the time-interval in the simulation. Since we used the log-log plots in Fig.~\ref{fig_dist2}, the bin size is too narrow in the right-hand region of the figures; there may be only one data point in a bin on the histogram. This causes the probability values to become too small. Therefore, we also colored densities with non-zero values using kernel density estimation. We here employed \texttt{scipy.stats.gaussian\_kde} for the kernel density estimation. The color of each point in the histogram was changed according to the probability density values obtained from kernel density estimation.

The distribution evaluated by the original trajectory, Fig.~\ref{fig_dist2}(a), shows a power-law decay and a peak structure away from the power-law. One might consider that the behavior in Fig.~\ref{fig_cross_correlation} is periodic, but there are many switchings with small time intervals, which lead to the power-law decay in Fig.~\ref{fig_dist2}(a). Note that we can see the power-law decay in all cases in Figs.~\ref{fig_dist2}(b), \ref{fig_dist2}(c) and \ref{fig_dist2}(d). However, the tendency to concentrate apart from the power-law decay is seen only in Figs.~\ref{fig_dist2}(a) and \ref{fig_dist2}(d). Although Fig.~\ref{fig_dist2}(d) may not look like a peak structure, the yellow area is away from the power-law decay, which corresponds to the peak structure in Fig.~\ref{fig_dist2}(a). The peak structure indicates the periodic behavior in Fig.~\ref{fig_cross_correlation}. Hence, the trajectories within the high-density cluster region and the high-pass filtered data lose this periodicity. By contrast, the low-pass filtered data remains appropriately preserved for this feature. We show some numerical results for other initial frequency difference $\Delta f_{\mathrm{ini}}$ in Appendix~A.

\renewcommand{\arraystretch}{1.3}
\begin{table*}[tb]
\centering
\caption{Summary of characteristics in the distribution.}
\label{tab_summary}
\begin{tabular}{c|cccc}
\hline
 &  Original & Part  & High-pass & Low-pass \\
\hline
\begin{minipage}{70mm}
Peak shift in the distribution of being leader?
\end{minipage}
& $\bigcirc$
& $\bigcirc$
& $\times$
& $\bigcirc$ \\
\begin{minipage}{70mm}
\vspace{2mm}
Power-law in the distribution of \\leader switching-time intervals?
\end{minipage} & $\bigcirc$ &  $\bigcirc$  & $\bigtriangleup$& $\bigcirc$ \\
\begin{minipage}{70mm}
\vspace{2mm}
Peak structure in the distribution of \\leader switching-time intervals? 
\end{minipage} & $\bigcirc$ & $\times$ & $\times$ & $\bigcirc$ \\
\hline
\end{tabular}
\end{table*}

Table~\ref{tab_summary} summarizes the characteristics of the distribution. Since the exponent of the power-law decay in Fig.~\ref{fig_dist2}(c) is slightly different from the others, we marked it as $\bigtriangleup$. From these discussions, we judge that the low-pass filtered data contains the essential characteristics of the original trajectories of the short-term cross-correlation $C_1(t)$ and $C_2(t)$.

Of course, there could be other important statistics.  Furthermore, while the peak shift structure in Fig.~5(d) matches that in Fig.~5(a), it exhibits a more complex shape. The key point here is not to perfectly reproduce the statistics, but to investigate what constitutes the essential characteristics of laser chaos in reinforcement learning tasks, as denoted above. Next, we will confirm whether the stochastic model constructed by the low-pass filtered data works well in the reinforcement learning task, i.e., the MAB problem.

\subsection{The multi-armed bandit problem}

\begin{figure}[tb]
\centering
\includegraphics[width=85mm]{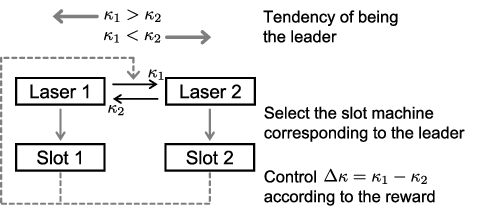}
\caption{Settings for the two-armed bandit problem with coupled lasers. A slot machine corresponding to the leader is selected, and the selected slot machine yields a positive (hit) or negative (miss) reward. The parameter $\Delta \kappa$ varies according to the reward, which controls the probability of being the leader.
}
\label{fig_slot}
\end{figure}

For simplicity, we here employ a similar experimental setting with \cite{Mihana2019}, i.e., the two-armed bandit problem. In \cite{Mihana2019}, the two coupled lasers were applied to the two-armed bandit problem. Different from the numerical experiments above, we employ $\kappa_1$ and $\kappa_2$ as the control parameters instead of the initial optical frequency detuning $\Delta f_{\mathrm{ini}}$. Then, we set $\Delta f_{\mathrm{ini}} = 0 \, \mathrm{Hz}$ and vary $\kappa_1$ and $\kappa_2$; the other parameter values are the same as in Table~\ref{tab_parameters}.

Since the experimental setting for the reinforcement learning is similar to that in \cite{Mihana2019}, we explain the key points briefly. Figure~\ref{fig_slot} shows the setting. The first and second lasers are assigned to two slot machines, respectively. When Laser 1 (or Laser 2) is the leader, the first (or second) slot machine is selected. The selected slot machine yields one of the results, hit or miss, and the result affects the control parameters $\kappa_1$ and $\kappa_2$, i.e., $\Delta \kappa \equiv \kappa_1 - \kappa_2$. The value of $\Delta\kappa(t)$ is updated as follows:
\begin{align}
\Delta\kappa(t) =
\begin{cases}
\,\, k N & \left(  \textrm{(int)}TA(t) > N\right), \\
\,\, k \textrm{(int)}TA(t) & \left( -N <  \textrm{(int)}TA(t) < N\right),\\
\,\, -k N & \left(   \textrm{(int)}TA(t)  < -N \right),
\end{cases}
\end{align}
where $k$ is a step width, and $TA(t)$ is a threshold adjuster value. The value of $TA(t)$ is updated as
\begin{align}
TA(t+1) = Z(t) + a \, TA(t)
\end{align}
where $a$ is a memory parameter, and $Z(t)$ is defined from the result of the slot machine. Let $\overline{P}_i$ be the estimated hit probabilities of the $i$-th slot machine, i.e., 
\begin{align}
\overline{P}_i = 
\frac{(\textrm{Number of hits for the $i$-th slot machine})}{(\textrm{Number of selections for the $i$-th slot machine})}.
\end{align}
Then, the value of $Z(t)$ is given as follows.
\begin{itemize}
\item First slot machine:
\begin{itemize}
\item hit: $Z(t) = 2-\overline{P}_1-\overline{P}_2$.
\item miss: $Z(t) = \overline{P}_1+\overline{P}_2$.
\end{itemize}
\item Second slot machine:
\begin{itemize}
\item hit: $Z(t) = -2+\overline{P}_1+\overline{P}_2$.
\item miss: $Z(t) = -\overline{P}_1-\overline{P}_2$.
\end{itemize}
\end{itemize}
The parameter values used in the numerical experiments are shown in Table~\ref{tab_multi_armed}. Using these experimental settings, we perform the exploration phase at the early stage. After achieving good estimates for the hit probabilities, the strategy is changed to the exploitation phase to obtain more rewards. These are the same as the previous studies, and then see the details in \cite{Mihana2019}.

\renewcommand{\arraystretch}{1.0}
\begin{table}[tb]
\centering
\caption{Parameter values used in the numerical experiments.}
\label{tab_multi_armed}
\begin{tabular}{c l l}
\hline
Symbol & Parameter & Value \\
\hline
$k$ &  Step width & $4$ [ns$^{-1}$]\\
$N$ & Number of step levels for positive $\Delta\kappa$ & $4$  \\
$\frac{2N}{k}+1$ & Total number of step levels & $9$\\
$a$ & Memory parameter & $0.999$\\
$P_1$ & Hit probability of slot machine $1$ & $0.6$\\
$P_2$ & Hit probability of slot machine $2$ & $0.4$\\
$\tau_{\mathrm{SI}}$ & Sampling interval & 7.0 [ns] \\
\hline
\end{tabular}
\end{table}

We here perform the computation $1000$ plays per cycle and repeat this for $100$ cycles ($n=100$) to compute the average over the $100$ runs and calculate average total rewards (ATR), which is defined as follows:
\begin{align}
\mathrm{ATR} = \frac{1}{n}\sum^n_{i=1}R_i,
\end{align}
where $R_i$ is the total reward over $1000$ plays.

As a result, the ATR values were $0.5700$ for the original deterministic laser chaos case and $0.5751$ for the constructed stochastic model; the estimated hit probabilities, $(\overline{p}_1,\overline{p}_2)$, were $(0.5999,0.3986)$ for the original deterministic laser chaos case and $(0.5624,0.3920)$ for the constructed stochastic model. Numerical results with different experimental settings are shown in Appendix~B.

In addition, we also examine the average convergence time (ACT); the ATR corresponds to the performance after learning, while the ACT relates to the dynamics of the learning process. Here, we define the ACT as follows: we evaluate the time at which learning converges, i.e., $\Delta \kappa(t) = 16$ or $-16$, and take the average of these times as the ACT. For the original deterministic laser chaos case, the ACT value is 112.53; in the case with the constructed stochastic model, we have 115.35 as the ACT value. Then, the constructed stochastic model yields a similar ACT value to that of the original laser chaos data.

From these results, we conclude that the constructed stochastic model achieves the same performance as the original deterministic laser chaos in the reinforcement learning problem.

\section{Conclusion}
\label{sec_conclusion}

In this study, we discussed the stochastic nature behind the deterministic laser chaos. Specifically, we attempted to reproduce the switching of the leader-laggard relationship observed in mutually coupled lasers with time delay. Using data preprocessing and the gEDMD algorithm, we explicitly constructed the stochastic models for the short-term cross-correlations given by the coupled lasers. The discussions clarified that at least low-pass filtering conserves the essential parts and that stochasticity, rather than deterministic chaos, could be enough for the final result of reinforcement learning.

As far as we know, the present work is the first trial to construct stochastic models for coarse-grained data generated by deterministic laser chaos. Since the methodology and discussions are the first step to investigating the essential parts of the deterministic laser chaos, further research is necessary. First, the stochastic model obtained by the current gEDMD algorithm is still difficult to interpret while it is useful for numerical simulations; the estimated Koopman generator matrix contains numerous terms, which prevents us from understanding the stochastic model intuitively. Then, further improvement of the gEDMD algorithm is desirable to extract the essence of the complex behavior of deterministic chaos. Second, there is the curse of dimensionality; as the dimension of the space increases, the size of the dictionary increases exponentially. As for this issue, approaches utilizing the tensor-train format have been proposed \cite{Klus2018}; applying the tensor-train format would be hopeful for the discussions of stochastic modeling in high-dimensional cases. Finally, we performed our numerical experiments as a demonstration. Hence, we did not perform fine-tunings. The optimal parameters vary depending on the model, although we determined them empirically. Therefore, the selection of hyperparameters could be an issue for future studies of model estimation. These future developments will contribute to understanding the nature of complex data, including deterministic chaos.

\begin{acknowledgments}
This work was financially supported in part by grants awarded to JO (JSPS KAKENHI Grant Numbers JP23H04800 and JP25H01880).
\end{acknowledgments}

\appendix

\section{Additional numerical results with different initial optical frequencies}

\begin{figure}[b]
\centering
\includegraphics[width=80mm]{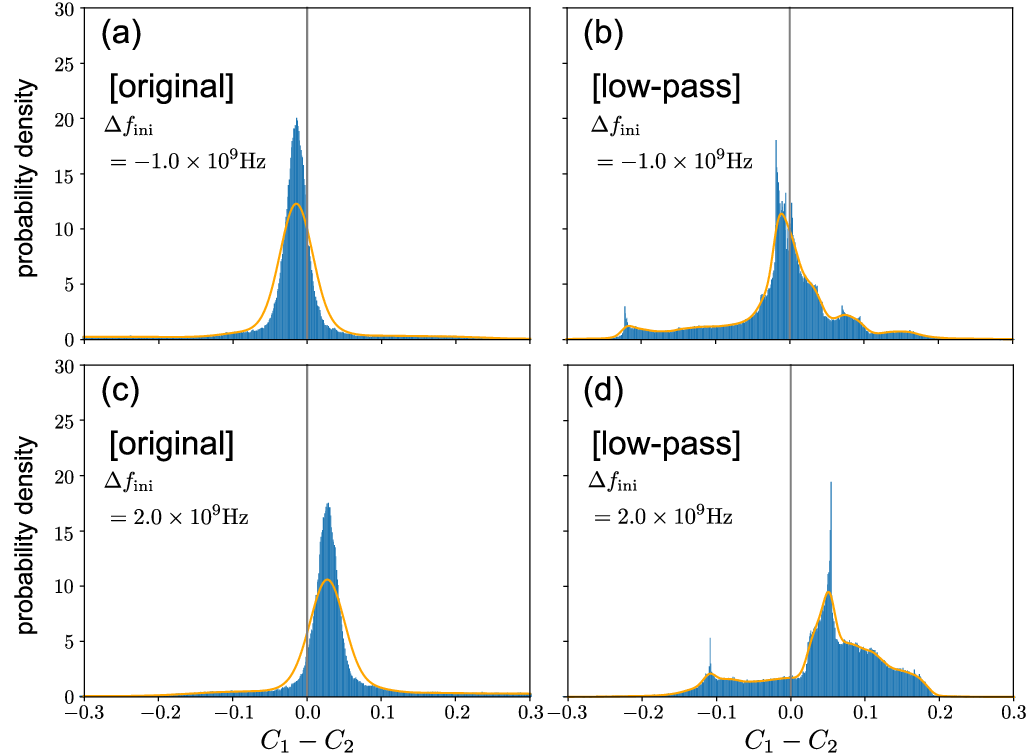}
\caption{The distribution of being the leader. For (a) and (b), $\Delta f_{\mathrm{ini}}=-1.0\times 10^{9}$ Hz are used; for (c) and (d), we used $\Delta f_{\mathrm{ini}}=2.0\times 10^{9}$ Hz. The other experimental settings are the same with Fig.~\ref{fig_dist1}.
}
\label{fig_distA1}
\end{figure}

In Figs.~\ref{fig_dist1} and \ref{fig_dist2}, we only showed the numerical results for $\Delta f_{\mathrm{ini}}$. In the MAB problem, the peak position must change in accordance with the initial frequency difference. We performed several numerical experiments with various settings and confirmed that the peak trend changes similarly to the original model. Figure~\ref{fig_distA1} shows the results for cases $\Delta f_{\mathrm{ini}}=-1.0\times 10^{9}$ Hz and $\Delta f_{\mathrm{ini}}=2.0\times 10^{9}$ Hz. These results indicate that a negative initial frequency produces a negative peak, and that the magnitude of the peak shift increases with the magnitude of the initial frequency.

\begin{figure*}[tb]
\centering
\includegraphics[width=120mm]{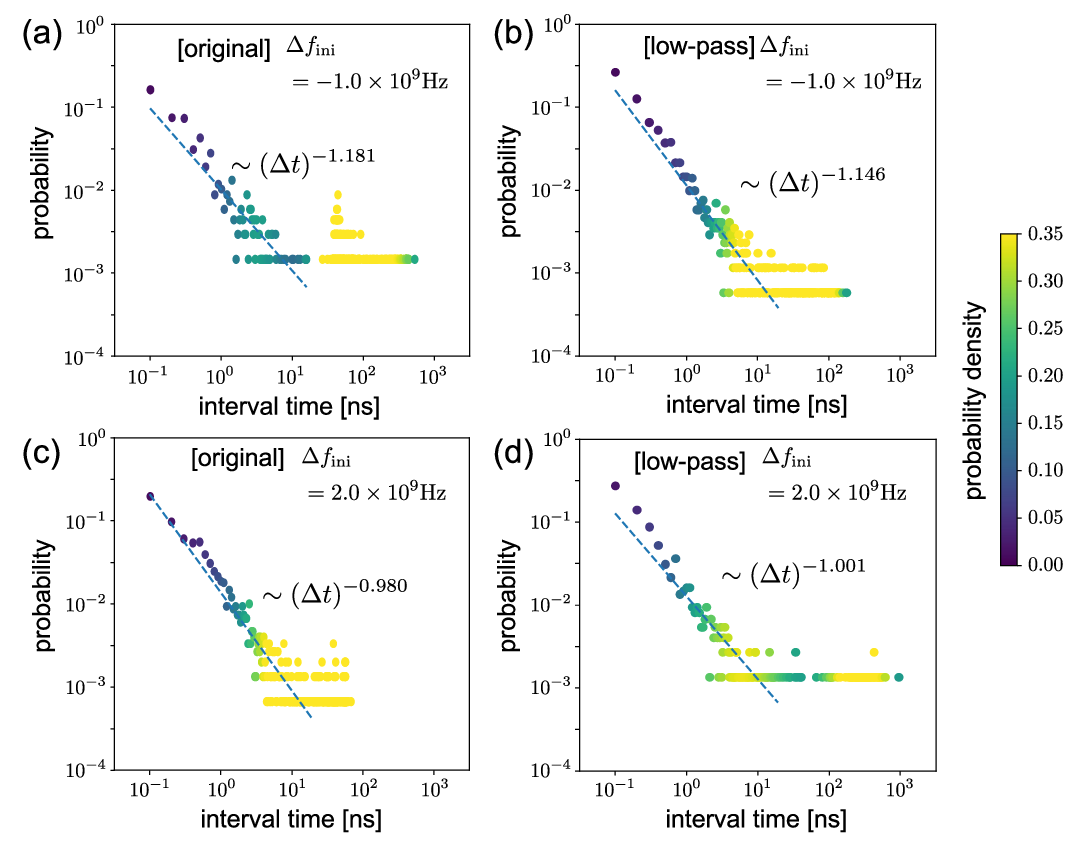}
\caption{The distribution of leader switching-time intervals from Laser 1 to Laser 2. 
For (a) and (b), $\Delta f_{\mathrm{ini}}=-1.0\times 10^{9}$ Hz are used; for (c) and (d), we used $\Delta f_{\mathrm{ini}}=2.0\times 10^{9}$ Hz. The other experimental settings are the same with Fig.~\ref{fig_dist2}.
}
\label{fig_distA2}
\end{figure*}

Figure~\ref{fig_distA2} shows numerical results for distribution of leader switching-time intervals. We confirmed that the stochastic model constructed with the low-pass filtered data recovers the essential features of the original data, as discussed in Sec.~IV.C.

\section{Additional numerical results for the MAB problem}

First, we show the MAB problem results for the other data preprocessing settings. In Sec.~IV.D, the hit probabilities are set to ($P_1, P_2$) are (0.6,0.4). When we employ the constructed model from the high-density cluster region datasets, the estimated hit probabilities ($\bar{p}_1,\bar{p}_2$) were (0.4687,0.1571), and the ATR value was 0.5532. As for the high-pass filtered case, we have ($\bar{p}_1,\bar{p}_2$) were (0.5997,0.3999), and the ATR value was 0.5083. Other numerical experiments also showed that the low-pass results were closest to the original ones. Hence, we conclude that the probabilistic model constructed from the low-pass data would capture the essential features for the MAB problems.

Second, we show the results with the different hit probability settings ($P_1, P_2$), i.e., (0.7,0.3) and (0.8,0.2). 

When ($P_1,P_2$)=(0.7,0.3), the ATR values were 0.6595 for the original deterministic laser chaos case and 0.6720 for the constructed stochastic model. The estimated hit probabilities, ($\bar{p}_1,\bar{p}_2$) were (0.7025,0.2985) and (0.6809,0.2827) for the original deterministic laser chaos case and for the constructed stochastic model, respectively.

When ($P_1,P_2$)=(0.8,0.2), the ATR values were 0.7419 and 0.7694, ($\bar{p}_1,\bar{p}_2$)=(0.8012,0.1987) and ($\bar{p}_1,\bar{p}_2$)=(0.7838,0.1997), for the original deterministic laser chaos case and for the constructed stochastic model, respectively.

\end{document}